\documentclass[5p,sort&compress]{elsarticle}

\usepackage{amsmath}
\usepackage{amssymb}
\usepackage{graphicx}
\usepackage{dcolumn}
\usepackage{bm}
\usepackage{mathtools}

\newcommand{\nuc}[2]{${}^{{#1}}\textrm{{#2}}$}
\newcommand{\nucs}[2]{${}^{{#1}}\textrm{{#2}}^*$}

\begin{document}

\begin{frontmatter}
  
\title{Single-neutron knockout from $^{20}\textrm{C}$ and the structure of $^{19}\textrm{C}$}



\author[snu]{J.W.~Hwang\corref{cor}} 
\ead{hjw8707@snu.ac.kr}
\author[snu]{S.~Kim}
\author[snu]{Y.~Satou}
\author[caen]{N.A.~Orr}
\author[titech]{Y.~Kondo}
\author[titech]{T.~Nakamura}
\author[caen]{J.~Gibelin}
\author[caen]{N.L.~Achouri}
\author[kern,emmi]{T.~Aumann}
\author[riken]{H.~Baba}
\author[caen]{F.~Delaunay}
\author[riken]{P.~Doornenbal}
\author[riken]{N.~Fukuda} 
\author[riken]{N.~Inabe}
\author[riken]{T.~Isobe}
\author[riken]{D.~Kameda}
\author[titech]{D.~Kanno}
\author[titech]{N.~Kobayashi}
\author[tohoku]{T.~Kobayashi}
\author[riken]{T.~Kubo}
\author[caen]{S.~Leblond}
\author[riken]{J.~Lee\fnref{preadd}}
\fntext[preadd]{Present Adress: Department of Physics, University of Hong Kong, Pokfulam Road, Hong Kong}
\author[caen]{F.M.~Marqu\'es}
\author[titech]{R.~Minakata}
\author[riken]{T.~Motobayashi}
\author[rikkyo]{D.~Murai}
\author[kyoto]{T.~Murakami}
\author[tohoku]{K.~Muto}
\author[kyoto]{T.~Nakashima}
\author[kyoto]{N.~Nakatsuka}
\author[ganil]{A.~Navin}
\author[titech]{S.~Nishi}
\author[titech]{S.~Ogoshi}
\author[riken]{H.~Otsu}
\author[riken]{H.~Sato}
\author[riken]{Y.~Shimizu}
\author[riken]{H.~Suzuki}
\author[tohoku]{K.~Takahashi}
\author[riken]{H.~Takeda}
\author[riken]{S.~Takeuchi}
\author[titech]{R.~Tanaka}
\author[emmi]{Y.~Togano}
\author[york]{A.G.~Tuff}
\author[paris]{M.~Vandebrouck}
\author[riken]{K.~Yoneda}

\cortext[cor]{Corresponding author}

\address[snu]{Department of Physics and Astronomy, Seoul National University,
1 Gwanak-ro, Gwanak-gu, Seoul 08826, Republic of Korea}
\address[caen]{LPC-Caen, IN2P3/CNRS, ENSICAEN, UNICAEN et Normandie Universit\'{e},
14050 Caen Cedex, France}
\address[titech]{Department of Physics, Tokyo Institute of Technology, 
2-12-1 O-Okayama, Meguro, Tokyo 152-8551, Japan}
\address[kern]{Institut f\"ur Kernphysik, Technische Universit\"{a}t Darmstadt, 
D-64289 Darmstadt, Germany}
\address[emmi]{ExtreMe Matter Institute EMMI and Research Division, 
GSI Helmholtzzentrum f\"ur Schwerionenforschung GmbH, D-64291 Darmstadt,
Germany}
\address[riken]{RIKEN Nishina Center, 
  Hirosawa 2-1, Wako, Saitama 351-0198, Japan}
\address[tohoku]{Department of Physics, Tohoku University, 
Miyagi 980-8578, Japan}
\address[rikkyo]{Department of Physics, Rikkyo University, 
Toshima, Tokyo 171-8501, Japan}
\address[kyoto]{Department of Physics, Kyoto University, 
  Kyoto 606-8502, Japan}
\address[ganil]{GANIL, CEA/DRF-CNRS/IN2P3, 
F-14076 Caen Cedex 5, France}
\address[york]{Department of Physics, University of York, 
Heslington, York YO10 5DD, United Kingdom}
\address[paris]{Institut de Physique Nucl\'{e}aire, Universit\'e
Paris-Sud, IN2P3/CNRS,  91406 Orsay, France}

\begin{abstract}

The low-lying unbound level structure of the halo nucleus \nuc{19}{C}
has been investigated using single-neutron knockout from \nuc{20}{C}
on a carbon target at 280~MeV/nucleon.  The invariant mass spectrum, derived from the momenta of the forward going beam 
velocity \nuc{18}{C} fragment and neutrons, was found to be dominated by a very narrow near threshold ($E_\textrm{rel} = 0.036(1)$~MeV) peak.
Two less strongly populated resonance-like features were also observed at $E_\textrm{rel} = 0.84(4)$ and 2.31(3)~MeV, both
of which exhibit characteristics consistent with neutron $p$-shell hole states. 
Comparisons of the energies, measured cross sections and
parallel momentum distributions to the results of shell-model and eikonal reaction calculations
lead to spin-parity assignments of 
$5/2^+_1$ and $1/2^-_1$ for the levels at $E_x = 0.62(9)$ and 2.89(10)~MeV with $S_n = 0.58(9)$~MeV.
Spectroscopic factors were also deduced and found to be in reasonable accord with shell-model calculations.
The valence neutron configuration of the $^{20}$C ground state is thus seen to include, in addition to the known $1s_{1/2}^2$ 
component, a significant $0d_{5/2}^2$ contribution.
The level scheme of \nuc{19}{C}, including significantly the $1/2^-_1$ cross-shell state, is
well accounted for by the YSOX shell-model interaction developed from the 
monopole-based universal interaction.  

\end{abstract}


\end{frontmatter}



The atomic nucleus is a finite fermionic quantum system that exhibits shell structure.
The manner and mechanisms by which this evolves with the neutron-proton $(N/Z)$ asymmetry
across the nuclear landscape is one of the key questions in nuclear structure physics.
Such investigations may be traced back to the early work of Talmi and Unna \cite{Talmi1960} where the ordering of the lowest-lying levels in \nuc{11}{Be} and \nuc{15}{C}
was discussed in terms of the residual shell-model interaction \cite{Talmi1962}.
Since these pioneering studies, the $p$-$sd$-shell nuclei have provided an important testing 
ground to explore our understanding of shell structure away from stability.  Experimentally, such studies are now possible beyond the proton and neutron driplines, as evidenced by recent measurements of
the most exotic oxygen isotopes \cite{Suzuki2009,Lunderberg2012,Kohley2013,Caesar2013,Kondo2016}.
Theoretically, the description of such near-drip-line nuclei is now possible using 
sophisticated models, ranging from the shell model to {\it ab initio} approaches, which include, explicitly or implicitly effects, such as three-nucleon forces, the continuum (and coupling to it for weakly bound levels), and tensor forces (see, for example, Refs.~\cite{Volya2005, Hagen2014, Jansen2014, Otsuka2010, Yuan2012}).

Of particular note in the context of the work presented here are shell-model calculations employing effective interactions derived from {\it ab initio} coupled-cluster (CCEI) theory which are now capable of predicting the binding
energies and low-lying levels for the most neutron-rich carbon and oxygen isotopes \cite{Jansen2014}.
In contrast, Otsuka {\it et al.}~have constructed a monopole-based universal interaction ($V_\textrm{MU}$)
consisting of the central and $\pi + \rho$ tensor terms \cite{Otsuka2010} which has provided
intriguing insight into changes in shell structure, including the neutron-rich $p$-$sd$-shell nuclei \cite{Yuan2012}.
This Letter reports the observation of $5/2^+$ and $1/2^-$ states in \nuc{19}{C}
populated via single-neutron knockout from \nuc{20}{C} at 280~MeV/nucleon.
The results are discussed in the context of a range of shell-model calculations, including those just mentioned,
and conclusions are drawn regarding the underlying shell structure.  Importantly, the observation of the $1/2^-$ 
neutron $p$-shell hole state provides a direct test of the cross-shell components of the shell-model interactions.

The nucleus $^{19}\textrm{C}$ is the heaviest bound odd-$A$ carbon isotope and 
the lightest member of the $N=13$ isotonic chain.  Structurally it  
is one of the few well established single-neutron halo nuclei \cite{Bazin1995,Marques1996,Baumann1998} 
with a very weakly bound $s$-wave valence neutron ($S_n=0.58(9)$~MeV \cite{Wang2012})
and ground state spin-parity $J^{\pi} = 1/2^+$ \cite{Nakamura1999,Maddalena2001}.
The low-lying level structure of $^{19}\textrm{C}$ is expected to be composed of
$1/2^+$, $3/2^+$, and $5/2^+$ states,
arising from neutron occupancy of the almost degenerate $0d_{5/2}$ and $1s_{1/2}$ orbitals \cite{Stanoiu2008}.
Although most shell-model predictions suggest that these states are closely spaced and 
located well below 1~MeV, their ordering has been the subject of considerable uncertainty 
including, in particular, the location of the $5/2_1^+$ level.
 
The first in-beam $\gamma$-ray spectroscopy of $^{19}\textrm{C}$ employed 
the $(p,p')$ reaction in inverse kinematics, 
and identified cascade transitions
consistent with two bound excited states at 0.196(6) and 0.269(8)~MeV \cite{Elekes2005}, 
which were tentatively assigned $3/2^+$ and $5/2^+$, respectively.
A measurement employing fragmentation of a mixed secondary
beam confirmed the existence of the transition
from the $3/2^+$ state to the ground state \cite{Stanoiu2008}.
A subsequent invariant mass study, also using the $(p,p')$ reaction in inverse kinematics, 
observed an unbound level at 1.46(10)~MeV, the angular distribution of which was consistent with a $5/2^+$ state \cite{Satou2008}. 
More recently, investigations of inclusive two-neutron removal from $^{20}\textrm{C}$ suggested, 
through comparison with eikonal reaction model calculations and shell-model spectroscopic factors,  
that the $5/2^+_1$ state should be unbound \cite{Ozawa2011,Kobayashi2012}, 
in contradiction with the conclusions of Ref.~\cite{Elekes2005}.  Subsequently
a candidate for the $5/2^+_1$ state was observed just above threshold ($E_x=0.693(95)$~MeV)
in the $^{18}\textrm{C}$ + neutron invariant mass spectrum following multi-nucleon removal 
from $^{22}\textrm{N}$ \cite{Thoennessen2013}\footnote{The tentative $5/2^+_1$ asignment was based on a comparison with shell-model excitation energies.}.

Recently two further in-beam $\gamma$-ray measurements were reported \cite{Whitmore2015,Vajta2015}.
Both confirmed the existence of a level, assigned $3/2^+$, at 0.20~MeV, whilst 
the former also provided a measure of the lifetime and $B(M1)$ strength.
In summary, the lowest two states -- the ground $1/2^+_1$ halo state and the $3/2^+_1$ level at 0.20~MeV --
are bound, whilst the $5/2^+_1$ state most probably lies just above  
the neutron decay threshold.  As will be discussed, the present work confirms this conjecture (and provides a clear
$d$-wave assignment)
and observes two more higher-lying resonances, one of which is identified as the lowest-lying 
negative parity state in \nuc{19}{C}.  

In terms of the \nuc{20}{C} projectile, the momentum distribution and the associated cross section 
for the C(\nuc{20}{C},\nuc{19}{C}) reaction, in the aforementioned inclusive neutron removal study \cite{Kobayashi2012}, 
reveal the presence of a significant $1s_{1/2}^2$ valence neutron configuration.  The expected $0d_{5/2}^2$ component was not
probed, as the corresponding $5/2^+$ level in \nuc{19}{C} is, as noted above, unbound.
It is worthwhile noting that the structure of \nuc{20}{C} is of interest, not only in terms
of shell evolution around the $N=14$ sub-shell closure \cite{Stanoiu2008}, but as the core of the heaviest two-neutron halo 
system \nuc{22}{C} \cite{Kucuk2014,Togano2016}. 


The experiment was performed at the Radioactive Isotope Beam Factory (RIBF) \cite{Yano2007} of the RIKEN Nishina Center
as a part of a series of measurements investigating
the structure of light neutron-rich nuclei
beyond the dripline (see, for example, Ref.~\cite{Kondo2016}).
A 345-MeV/nucleon $^{48}\textrm{Ca}$ primary beam ($\sim$ 100~pnA) 
impinging on a 20-mm-thick Be production target was employed, in conjunction with the 
BigRIPS separator \cite{Kubo2003},
to produce a mixed secondary beam, including $^{20}\textrm{C}$ at an average rate of 190~pps.
The various isotopes present in the secondary beam were identified event-by-event using measurements of 
the energy loss, time-of-flight, and magnetic rigidity.
The secondary beam was transported to the object point of the SAMURAI spectrometer \cite{Kobayashi2013}
where a carbon reaction target
with a thickness of 1.8~g/cm$^2$ was located.
The beam particles were tracked onto the target using two drift chambers.
The $^{20}\textrm{C}$ mid-target energy was 280~MeV/nucleon.  Data were also acquired with the 
carbon target removed in order to account for reactions in the various beam detectors.

The forward-focused beam velocity reaction products, including \nuc{18}{C} and a neutron, were detected 
using the SAMURAI spectrometer and large area NEBULA neutron array \cite{Nakamura2015}.
The charged fragments
were momentum analyzed by the 3~T superconducting dipole magnet,
and the magnetic rigidity deduced using the trajectories derived from drift chambers
placed at the entrance and exit of the magnet as described in Ref.~\cite{Kobayashi2013}.
A 16-element plastic hodoscope provided for energy loss and time-of-flight measurements, which combined with the rigidity permitted the charged fragments to be identified.

The NEBULA array was located some 11~m
downstream of the secondary target. The array comprised 120 individual
detector modules (each 12~cm $\times$ 12~cm $\times$ 180~cm)
and 24 charged particle veto detectors (thickness 1 cm),
arranged in a two-wall configuration, with an interwall
separation of 85~cm.  The neutron momenta were derived from the time-of-flight (measured with respect to a plastic detector placed forward of the secondary target) and hit position.  

The $\gamma$ rays emitted from excited states of the charged fragments
were detected using 140 NaI(Tl) scintillators of the DALI2 array \cite{Takeuchi2014} which were arranged in a 4$\pi$-like configuration around the secondary reaction target.  As such, the array had a detection efficiency of 
16\% at 1~MeV and an energy resolution (FWHM) of 150~keV.

The relative energy ($E_{\text{rel}}$)
of $^{19}\textrm{C}^*$ was reconstructed from the four-momenta of the 
$^{18}\textrm{C}$ fragment and decay neutron.
Specifically, the $E_{\text{rel}}$ was calculated as,
\begin{equation}
E_{\text{rel}} = \sqrt{ (E_f + E_n )^2 - | \bm{p}_f + \bm{p}_n |^2 } - (M_f + M_n),
\end{equation}
where $E_f$ ($E_n$), $\bm{p}_f$ ($\bm{p}_n$), and $M_f$ ($M_n$) 
are the total energy, momentum, and mass of $^{18}\textrm{C}$ (neutron), respectively.

In the eikonal-model description of nucleon knockout, neutrons are removed from the 
$^{20}\textrm{C}$ projectile via absorption and diffraction \cite{Tostevin1999}.  At the 
present beam energies the former process dominates.
The small fraction ($\sim10\%$) of diffractive breakup events are associated with two beam-velocity neutrons
in the outgoing channel in coincidence with $^{18}\textrm{C}$.  As such, a very broad low-level background \cite{Pain2006}, in addition to the $^{19}\textrm{C}^*$ continuum, is expected (as verified by simulations) in the $E_{\text{rel}}$ spectrum.    

The longitudinal momentum ($p_\parallel$) of \nucs{19}{C} 
was deduced from the sum of $\bm{p}_f$ and $\bm{p}_n$ after correcting for the spread in $^{20}\textrm{C}$ beam momenta.  The $p_\parallel$ and $E_{\text{rel}}$ distributions shown in the following were obtained after subtracting the 
contributions arising from material other than the secondary reaction target.


\begin{figure}[!t]
\centering
\includegraphics[width=0.5\textwidth]{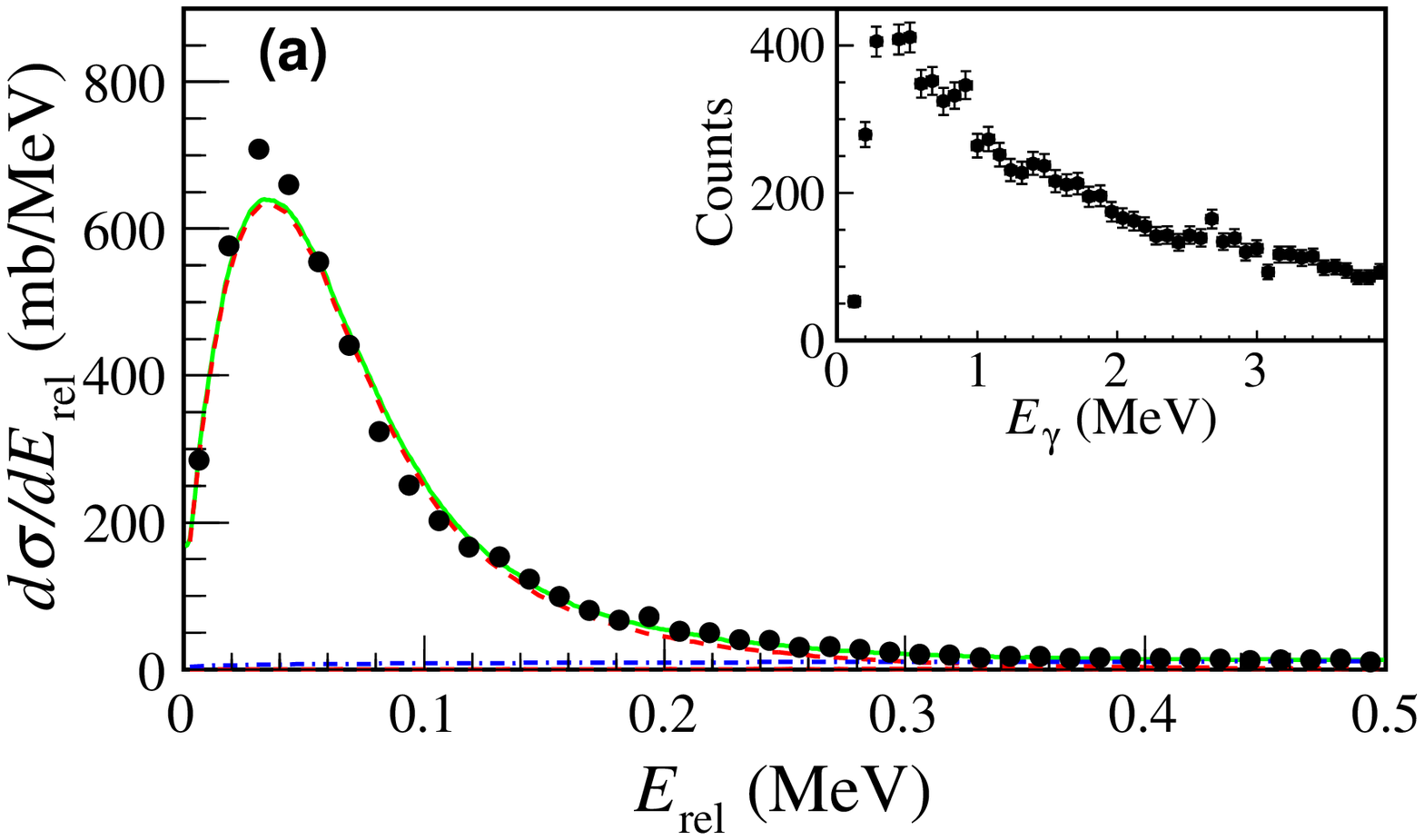}
\includegraphics[width=0.5\textwidth]{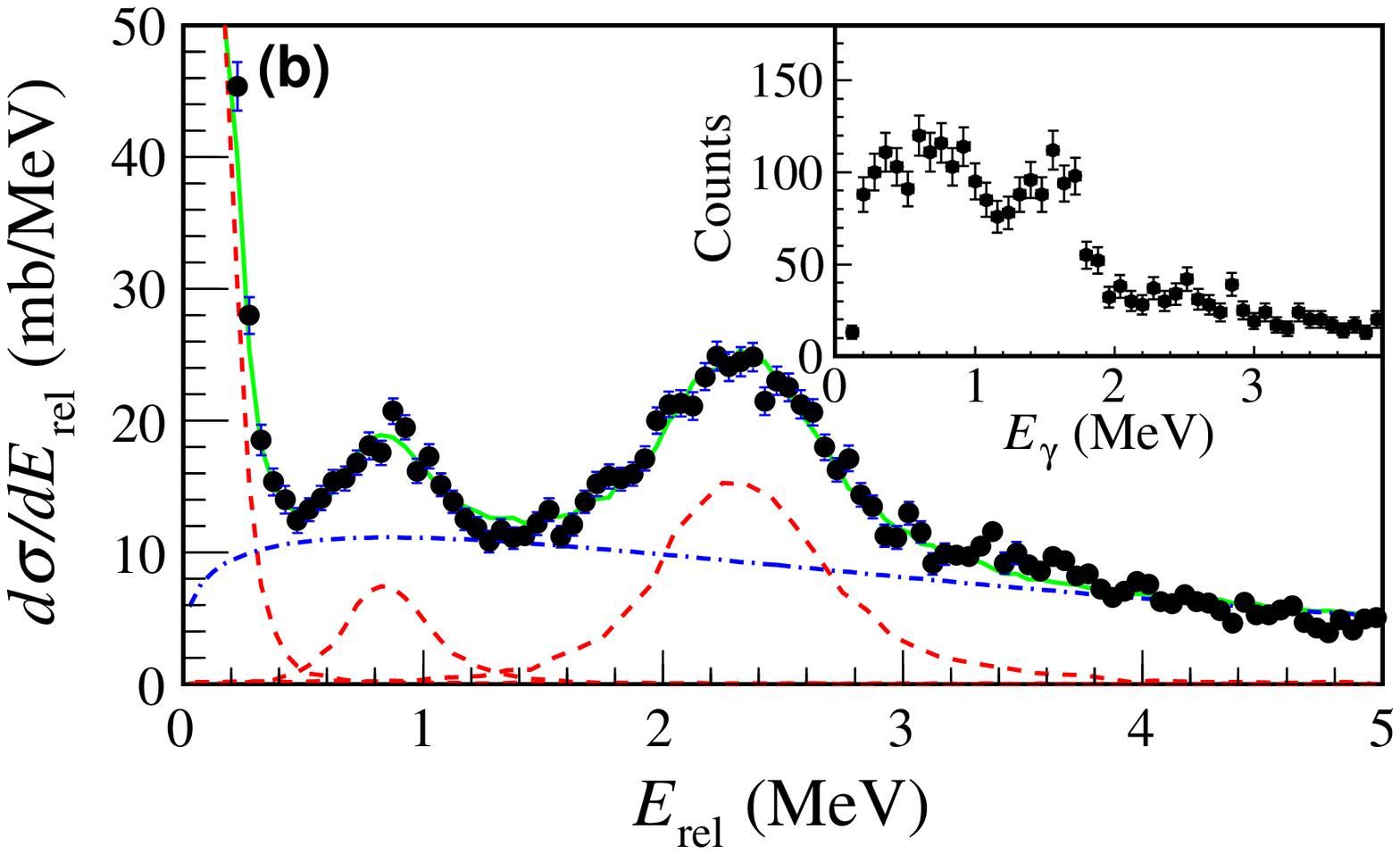}
\caption{(Color online) Relative energy spectrum for the $^{18}\textrm{C}+n$ system (solid points with error bars) up to (a) 0.5~MeV and (b) 5~MeV.
The solid (green) curve shows the results of the fit to the overall spectrum. 
The dashed (red) and dot-dashed (blue) curves represent the lineshapes of the individual 
resonances and background, respectively. 
The inset of panel (a) displays 
the Doppler-corrected energy spectrum of $\gamma$ rays 
in coincidence with the threshold peak -- $E_{\textrm{rel}} < 0.2$~MeV -- while that of panel (b) the spectrum in
coinicidence with events in the range $E_{\textrm{rel}} = 0.5$--1.3~MeV.}
\label{fig:erel}
\end{figure}

The $^{18}\textrm{C}+n$ $E_{\textrm{rel}}$ spectrum (Fig.~\ref{fig:erel}) exhibits a very prominent 
narrow threshold peak together with two more weakly populated higher-lying structures. 
In order to display the results in terms of the differential cross section, $d\sigma/dE_{\textrm{rel}}$, the geometrical acceptances and detection efficiencies have been taken into account. 
The former were evaluated, as a function of $E_{\textrm{rel}}$, using a complete simulation of the setup, which included the characteristics of the 
$^{20}\textrm{C}$ secondary beam and the momentum imparted to $^{19}\textrm{C}^*$ by the knocked-out neutron.  

In order to describe qualitatively the $E_{\text{rel}}$ spectrum, three single-level R-matrix lineshapes \cite{Lane1958}, convoluted with the experimental resolution function, and a very broad distribution (representing the continuum and diffracted neutron background -- see above) were employed, following
similar procedures to those detailed in Ref.~\cite{Satou2008}.
The resolution function, generated by simulations incorporating the effects of all the relevant detectors\footnote{The NEBULA hit position and timing resolutions being the dominate contributions.},
varied as (FWHM) $\Delta E_{\text{rel}} \approx 0.40 \sqrt{E_{\text{rel}}}$~MeV.
The underlying continuum background distribution was modeled, in line with earlier work (see, for example, Ref.~\cite{Satou2014}), with a Maxwellian-like distribution with a functional form of $a \sqrt[4]{x} e^{-bx}$, where $x=E_{\text{rel}}$, and $a$ and $b$ were the fitting parameters.  It may be noted that the form of the continuum is rather strongly constrained by the minima at 0.5 and 1.4~MeV, the spectrum at high $E_{\textrm{rel}}$, and that the intensity at 0~MeV must be zero.

\begin{table*}[!t] 
\caption{Cross sections ($\sigma_{-1n}$) and excitation energies ($E_x$) of the unbound states in   
  $^{19}\textrm{C}$ produced via single-neutron knockout from $^{20}\textrm{C}$ compared with reaction and shell-model (WBP interaction \cite{Warburton1992}) calculations.
See text for discussion of the character of the peak at $E_{\textrm{rel}} = 0.84$~MeV.} 
\begin{tabular}{llllllllll}
\hline 
$E_{\text{rel}}$ (MeV) & $E_x$ (MeV) & $\mathit{\Gamma}$ (MeV) & $\ell$ ($\hbar$) & $\sigma^{\text{exp}}_{-1n}$ (mb) & $\sigma_{\textrm{sp}}$ (mb)$^{~a,b)}$ & $C^2S^{\text{exp}}$$^{~a)}$ & $C^2S^{\text{th}}$ & $E_x^{\textrm{th}}$ (MeV) & $J^{\pi}$ \\
\hline 
0.036(1) &  0.62(9) & $<$ 0.015  & 2 & 61(5) & 22.9 & 2.40(20) & 3.80 &  0.240 & $5/2^+_1$\\
0.84(4) & 3.0 -- 5.5$^{~c)}$ & $<$ 0.02   & 1 & 4(1) &   &   &  &   &  \\
2.31(3) &  2.89(10) & 0.20(7)  & 1 & 15(3) & 18.6 & 0.77(15) & 1.38 & 1.907 & $1/2^-_1$ \\ 
\hline
\label{tab:fit}
\end{tabular}
\begin{footnotesize}
\begin{tabular}{rl}
$^{a)}$ & An uncertainty, not tabulated, associated with the reaction modeling of $\pm 15\%$ is estimated for $\sigma_\textrm{sp}$ and hence $C^2S^\textrm{exp}$ (see text). \\
$^{b)}$ & $S_n^\textrm{eff}$ derived from the experimental $E_x$ were employed in the reaction calculations. \\
$^{c)}$ & See text. \\
\end{tabular}
\end{footnotesize}
\end{table*}

Resonance energies of 0.036(1), 0.84(4), and 2.31(3)~MeV were deduced,
where single-level R-matrix lineshapes \cite{Satou2008} with $\ell_n = 1$ and 2 dependencies, according to the spin-parity assignments made below, were employed.  In the case of the lowest two peaks the widths were dominated by the experimental resolution and only upper limits could be determined (Table \ref{tab:fit}).
As no obvious coincident $\gamma$ rays were observed
for the $^{18}\textrm{C}+n$ events\footnote{$E_x$($2^+_1$) = 1.6~MeV
in $^{18}\textrm{C}$ \cite{Stanoiu2008,Fifield1982}.} 
forming the near threshold and highest-lying peaks
(the inset of Fig.~\ref{fig:erel}(a) illustrates this
for the threshold state)
corresponding excitation energies in $^{19}\textrm{C}$ of 
0.62(9) and 2.89(10)~MeV,
where the uncertainty in $S_n (^{19}\textrm{C})$
has been included, were deduced.

In the case of the most weakly populated peak at $E_{\text{rel}}$ = 0.84~MeV, the coincident $\gamma$-ray 
spectrum (inset of Fig.~\ref{fig:erel}(b))
shows evidence for the feeding of the $^{18}\textrm{C}$($2^+_1$) state.  
Taking into account the detection efficencies and assuming that all of the 
observed 1.6-MeV $\gamma$ rays are associated with the $E_{\text{rel}} = 0.84$-MeV peak and not 
the underlying continuum, a branching ratio of order 100\% is deduced.  This 
suggests that a higher-lying level is being populated.  We return to the origin of this peak below.

Theoretically single-neutron removal cross section $\sigma_{-1n}$ leading to a given final state
can be expressed in a factorized form as \cite{Hansen2003},
\begin{equation}
\sigma_{-1n} = \sum_{n\ell j} \left(\frac{A}{A-1}\right)^N C^2S(J^{\pi},n\ell j) \sigma_{\text{sp}}(n\ell j,S_n^{\text{eff}}),
\end{equation}
where $\sigma_{\mathrm{sp}}$ is the single-particle cross section, $n \ell j$ denote the quantum numbers of the knocked-out neutron, $[A/(A-1)]^N$ is the center-of-mass correction factor
with $A$ the mass number of the projectile and $N$ the principal oscillator quantum number ($N = 2n + \ell$) \cite{Dieperink1974},
and $S_n^{\text{eff}}$ the effective one-neutron separation energy
given by the sum of $S_n$ of the projectile
($S_n (^{20}\textrm{C}) = 2.93(26)$~MeV \cite{Wang2012})
and $E_x$ of the state in question.
  
Shell-model spectroscopic factors ($C^2S$) were computed using the $\textsc{NuShellX@MSU}$ \cite{Brown2014} 
code and the WBP interaction \cite{Warburton1992}\footnote{Only small variations were found between the results
obtained using the WBP, WBT, and YSOX interactions.} in the $0p$-$1s0d$ model space (Table~1).
The $\sigma_{\text{sp}}$ and associated momentum distributions were computed using 
the MOMDIS code \cite{Bertulani2006}. 
The valence neutron wave function was calculated using a Woods-Saxon potential
and the well-depth prescription of Ref.~\cite{Gade2008}. 
The range parameter of the nucleon-nucleon profile function \cite{Ray1979} 
at the present energy (280~MeV/nucleon) was set 
to zero \cite{Hansen2003}. 

The nucleon density distribution of the $^{19}\textrm{C}$ core was estimated
from a Hartree-Fock calculation using the SkX interaction \cite{Brown1998}.
The density distribution of the carbon target
was chosen to be of a Gaussian form 
with a point-nucleon rms radius of 2.32~fm. 
An overall uncertainty, not included in the tabulated values, of $\pm 15 \%$ was assigned to $\sigma_\textrm{sp}$,
comprising $\pm 10 \%$ associated with uncertainties in the size of the unbound core
(corresponding changes of the core radius of $\pm 5 \%$)
and $\pm 10 \%$ arising from uncertainties in the reaction theory \cite{Sauvan2004,Carstoiu2004}.

\begin{figure}[!t]
\centering
\includegraphics[width=0.5\textwidth]{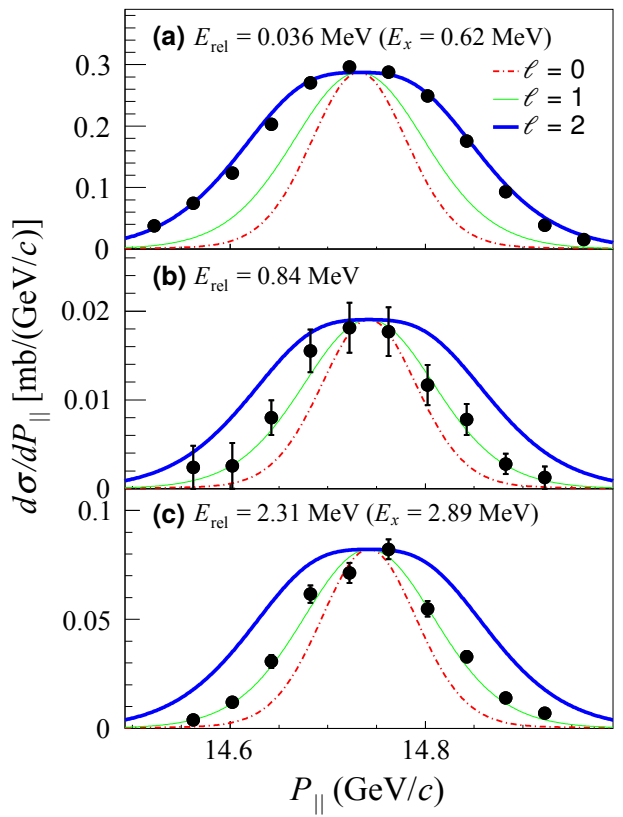}
\caption{(Color online) Experimental longitudinal momentum distributions (solid points)
compared with those computed for removal of neutrons with $\ell=0$, 1, and 2
(red dot-dashed, green solid, and thick blue solid lines, respectively)
for the states
at $E_x = 0.62$ (a) and 2.89~MeV (c) and the peak at $E_\textrm{rel} = 0.84$~MeV (b).  
The theoretical lineshapes have been
convoluted with the experimental resolution and, for the purpose of comparison, the
lineshapes are normalised to that which best fits the measurents (see text).}
\label{fig:momdis}
\end{figure}

Figure \ref{fig:momdis} shows the $^{19}\textrm{C}^*$ $p_{\parallel}$ distributions in the laboratory frame, after account was 
taken for the underlying continuum background, 
for the well defined levels at $E_x = 0.62$ and 2.89~MeV together with the peak
at $E_\textrm{rel} = 0.84$~MeV.
More specifically, for each momentum bin, the $E_{\text{rel}}$ spectrum
was fit assuming the three peaks and the continuum background distribution.
The error bars shown are statistical and the choice of the exact form for the continuum distribution did not change perceptibly
the form of the extracted momentum distributions. 
The experimental distributions are compared in each case in Fig.~\ref{fig:momdis} with the theoretical lineshapes,
convoluted with the experimental resolution ($\sigma \approx 28$~MeV/$c$ in the beam rest frame), for removal of
neutrons with orbital angular momentum $\ell = 0$, 1, and 2.
In the case of the $E_x = 0.62$-MeV state, the data are very well described when the removed neutron is of $d$-wave character.
The experimental distribution for the 2.89-MeV level 
is in very good agreement with removal of a $p$-wave neutron.

For the peak at $E_\textrm{rel} = 0.84$~MeV,
the $p_\parallel$ distribution is well reproduced assuming $p$-wave neutron removal ($\chi^2/n$=0.3, 3.0 and 5.7 for $\ell$=1, 0, and 2 respectively).
Interestingly, the apparent excitation energy assuming no feeding of the $^{18}\textrm{C}$($2^+_1$) state is
1.42(10)~MeV, very close to that of the 5/2$^+$ level observed ($E_x = 1.46(10)$~MeV) in the $(p,p')$ investigation \cite{Satou2008}, which,
based on the WBP interaction spectroscopic factor and eikonal model,
would be expected to be weakly populated ($\sim3$~mb).
The incompatibility of the momentum distribution with $d$-wave neutron removal is consistent, however, with the 
suggestion derived from the $\gamma$-ray coincidences (see above) that this peak arises from population of a higher-lying level in 
$^{19}\textrm{C}$ which has a decay branch that proceeds via the $^{18}\textrm{C}$($2^+_1$) excited state, rather than through neutron emission directly to the ground state.  It may also be 
noted that the neutron-decay width observed here ($\Gamma < 0.02$~MeV) 
is significantly smaller than in the inelastic scattering study \cite{Satou2008}.

Table \ref{tab:fit} summarizes the results where the uncertainties quoted for $E_x$ are dominated by the uncertainty in 
$S_n(^{19}\textrm{C})$.
Those assigned to the cross sections ($\sigma^{\text{exp}}_{-1n}$) arise from the uncertainty in the exact form 
for the continuum background distribution (5\%, 11\%, and 17\%
for the $E_{\textrm{rel}} = 0.036$, 0.84, and 2.31-MeV resonances, respectively),
the statistical uncertainty (2.5\%, 8.3\%, and 4.5\%), the 
neutron detection efficiency (5\% for all resonances),
and geometrical acceptance (2\%).

The energy of the state at $E_x = 0.62$~MeV
is consistent with that reported
by the multi-nucleon removal study of Ref.~\cite{Thoennessen2013}. 
The clear $\ell=2$ character of the momentum distribution and the large spectroscopic factor allow the state to be assigned
as the $5/2^+_1$ with good confidence -- the spectroscopic strength to $3/2^+$ levels is, unsurprisingly, expected 
to be very low ($C^2S \lesssim 0.25$).  The strong population of this level reflects the significant
$0d_{5/2}^2$ valence neutron configuration in $^{20}$C whereby the occupancy of the $0d_{5/2}$ neutron orbital is predicted 
to be around 4.3 \footnote{That of the $0d_{3/2}$ neutron orbital is predicted to be close to 0.5 and
1.3 for the $1s_{1/2}$ orbit.}.
It may also be noted that the unbound character of the $5/2^+_1$ level is in line with the earlier suggestions
of Refs.~\cite{Ozawa2011,Kobayashi2012,Thoennessen2013,Vajta2015}. 

\begin{figure}[!t]
\centering
\includegraphics[width=0.4\textwidth]{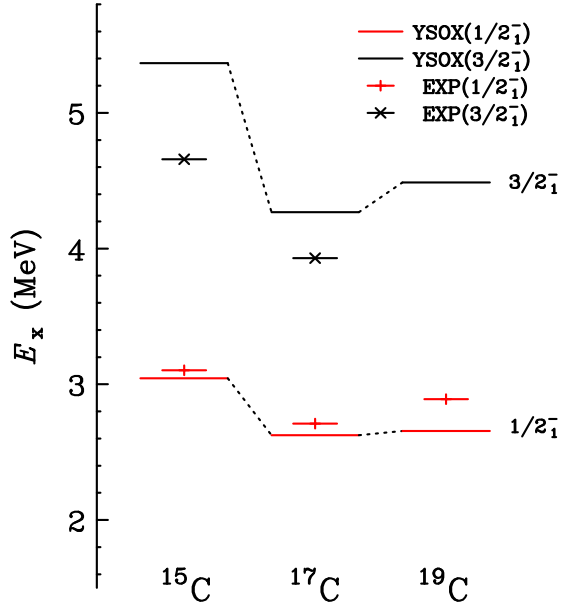}
\caption{(Color online) Excitation energies of the $1/2^-_1$ (red) and $3/2^-_1$ (black) levels in \nuc{15,17}{C} \cite{Ajzenberg1991,Harkewicz1991,Ueno2013} 
compared with shell-model calculations employing the YSOX interaction (see text).  In the case of 
\nuc{19}{C} the $\ell = 1$ resonance observed here at 2.89~MeV is displayed together with the shell-model predictions.}
\label{fig:systematics}
\end{figure}

The clear $\ell=1$ character of the momentum distribution associated with the
2.89-MeV level indicates a spin-parity of $1/2^-$ or $3/2^-$.  The moderate spectroscopic strength favours 
the $1/2^-$ assignment, which is reinforced by the location of the corresponding levels in 
\nuc{15,17}{C} \cite{Ajzenberg1991,Harkewicz1991,Ueno2013}.  As may be seen in Fig.~\ref{fig:systematics}, in both cases the $1/2^-_1$
state lies over 1~MeV below the $3/2^-_1$.  In addition, the YSOX interaction (see below), which
predicts very well the position of the $1/2^-_1$ level in \nuc{15,17}{C}, indicates it should lie in \nuc{19}{C} 
very close to the energy observed here and, once again, well below the $3/2^-_1$.

In the case of the $E_\textrm{rel} = 0.84$-MeV peak, the $\ell=1$ character of the associated momentum distribution and the
energy difference of 1.47(5)~MeV with respect to the relatively broad ($\mathit{\Gamma}=0.20(7)$~MeV) $1/2^-$ level suggest that
it could, in principle, arise from decay of the latter to the $^{18}\textrm{C}$($2^+_1$) state.
Shell-model calculations indicate, however, that the branching ratio for such a decay is negligible
and that the decay of the $1/2^-$ level proceeds essentially exclusively to the $^{18}\textrm{C}$ 
ground state\footnote{Note that if such a 
scenario were the origin of the $E_\textrm{rel} = 0.84$-MeV peak, the increase in yield to the $E_x = 2.89$-MeV level would be similar to the experimental uncertainty (Table~1) and, in terms of the spectroscopic factor, 
much smaller than the uncertainty ascribed to the reaction modeling.}.

\begin{figure}[!t]
\centering
\includegraphics[width=0.45\textwidth]{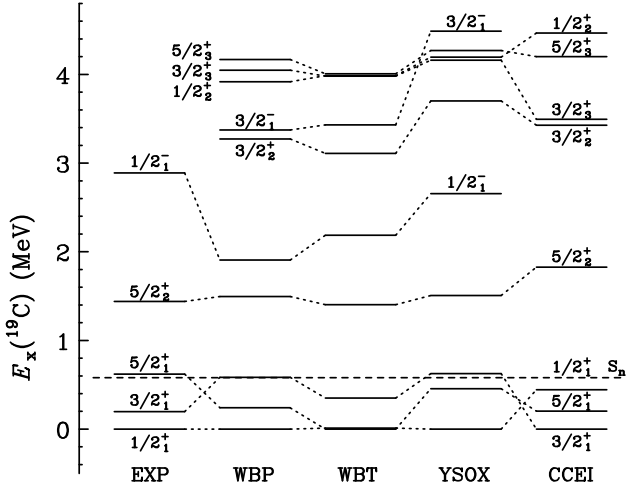}
\caption{Energies of states observed in $^{19}\textrm{C}$ (EXP: present work and Refs.~\cite{Elekes2005,Stanoiu2008,Satou2008,Thoennessen2013}) 
as compared to shell-model predictions ($E_\textrm{x} < 5$~MeV) 
for states with $J^{\pi} \leq 5/2^+$ and $3/2^-$ 
using the WBP, WBT \cite{Warburton1992}, YSOX \cite{Yuan2012},
and CCEI \cite{Jansen2014} interactions.
The latter are confined to $1s0d$-shell states only.}
\label{fig:level}
\end{figure}

The shell-model predictions (Fig.~\ref{fig:level}) place 
the first $3/2^-$ state above $\sim 3.0$~MeV excitation energy.  In terms of strength, the eikonal-model calculations suggest the 
cross section to be around half of that predicted for the population of the $1/2^-_1$ level.   
While the $3/2^-_1 $ state is calculated to have a reasonably strong
decay branch to the $^{18}\textrm{C}$($2^+_1$) level, placing it at $E_\textrm{x} = 3.02$~MeV, it is highly unlikely 
(see above and Fig.~\ref{fig:systematics}) that it is almost degenerate with the $1/2^-_1$ level.

Given then that the $3/2^-_1$ state almost certainly lies above the $1/2^-_1$, 
it is possible that the $E_\textrm{rel} = 0.84$-MeV peak 
could arise from decay to the $(0,2)^+$ level(s) at 2.5~MeV in $^{18}\textrm{C}$ \cite{Stanoiu2008,Kondo2009}, with a corresponding
excitation energy in  $^{19}\textrm{C}$ of 3.92~MeV.  
While the shell-model calculations suggest that
a reasonably strong decay branch to the $^{18}\textrm{C}$($2^+_2$) is possible, there is no clear sign of the 
corrsponding 0.92-MeV $\gamma$-ray transistion to the $^{18}\textrm{C}$($2^+_1$) state (inset Fig.~\ref{fig:erel}(b)),
nor the neutron decays of comparable strength predicted to $^{18}\textrm{C}$($0^+_1$) and ($2^+_1$) -- $E_\textrm{rel}$ = 3.34 and 1.74~MeV, respectively. 

The only other bound state(s) known in $^{18}\textrm{C}$ ($S_n =4.18(3)$~MeV \cite{Wang2012})
lies at 4.0~MeV with a probable $(2,3)^+$ assignment \cite{Stanoiu2008,Kondo2009}.  The shell model
suggests that decay to this level(s) may occur and would place the $3/2^-_1$ state at 5.42~MeV.
In this case the
2.4-MeV $\gamma$-ray transition to the $^{18}\textrm{C}$($2^+_1$) state could be difficult to identify owing to 
the detection efficiency.  In addition, the direct neutron decay branch to the $^{18}\textrm{C}$
ground state would be very difficult to observe owing to the low detection efficency and poor resolution
at high $E_\textrm{rel}$.  Such a scenario is, however, complicated by the two-neutron decay to $^{17}\textrm{C}$
being also energetically possible by 0.66~MeV.

It is clear that a more detailed investigation with a higher statistics data set is desirable.  While it is
not possible to provide a definitive conclusion, it is probable that the $E_\textrm{rel} = 0.84$-MeV peak 
arises from the neutron decay of the $3/2^-_1$ level 
to a bound excited state of $^{18}\textrm{C}$.  As such, the $3/2^-_1$ state may be expected to lie between  
3~MeV and 5.5~MeV excitation energy in $^{19}\textrm{C}$.  

Figure \ref{fig:level} displays a comparison of the energies of states observed in $^{19}\textrm{C}$
(present work and Refs.~\cite{Elekes2005,Stanoiu2008,Satou2008,Thoennessen2013}) with a range of different shell-model predictions.
All of the calculations were, except those labeled CCEI, performed using the $\textsc{NuShellX@MSU}$ code.
Results are shown 
for the WBP, WBT \cite{Warburton1992}, and YSOX \cite{Yuan2012} interactions in the $p$-$sd$ model space.
The results of calculations performed within the $sd$ shell-model space utilizing the {\it ab initio} 
Coupled-Cluster Effective Interaction (CCEI) are also shown
~\cite{Jansen2014}.  
In the case of the YSOX interaction, the $p$-$sd$ cross-shell components of the effective interaction
were constructed based on $V_\textrm{MU}$ \cite{Otsuka2010},
which was developed from data obtained closer to stability.
The CCEI interaction includes explicitly the effects of three-body forces 
derived from chiral effective-field theory.

While all of the models predict the occurrence of three very low-lying positive parity states
($1/2^+$, $3/2^+$, and $5/2^+$) none is able to reproduce the ordering.
Interestingly, although the CCEI shell-model calculations predict the ordering of the $3/2^+$ and $5/2^+$ levels,
the $1/2^+$ state is found to lie above both of them. 
However, as noted by Jansen {\it et al.}~\cite{Jansen2014}, the very weakly bound $s$-wave character of the 
$1/2^+$ state means that the effects of the 
coupling to the continuum need to be properly included.  Indeed, initial estimates suggest that after doing so the $1/2^+$
level is expected to be lowered, relative to the $3/2^+$ and $5/2^+$ states, by around 1~MeV.
It is worthwhile noting that the spacing between the $1/2_1^+$ and $5/2_1^+$ states reflects the behaviour of the
corresponding neutron single-particle orbits, which are, as noted earlier, expected to be almost 
degenerate in the very neutron-rich carbon isotopes \cite{Stanoiu2008}. 

The newly observed $1/2^-$ state at 2.89~MeV is best accounted for by the calculations employing the YSOX interaction.
This may be attributable to the cross-shell parts of the interaction incorporating $V_\textrm{MU}$.
Such an ability to describe neutron cross-shell states
in neutron-rich nuclei has also been noted in terms of the role of microscopic three-body forces,
for the $V_\textrm{MU}$-based shell-model interaction SDPF-MU \cite{Utsuno2012},
which was constructed in the $sd$-$pf$ model space and used to investigate the spectroscopy of $^{35,37,39}\textrm{Si}$~\cite{Stroberg2015}.
 
Recently, Hoffman {\it et al.}~\cite{Hoffman2014} have discussed the behavior of neutron $s$-wave states
in the context of finite binding effects which become significant for shallow binding.
The present study provides a measure of the relative $1/2^+$--$5/2^+$ separation in \nuc{19}{C}
of $-0.62(9)$~MeV which is close to that expected on the basis of the systematics
(See Fig.~4 (a) of Ref.~\cite{Hoffman2014}).
This behavior may also be seen in the manner in which the energy of the $1/2^+$ level drops relative to that of 
the $5/2^+$ level in the carbon isotopes as compared to the corresponding oxygen isotones.  Specifically, the $1/2^+$--$5/2^+$ separation is reduced, by an almost constant amount, for the $N = 9$, 11, and 13 isotones:
$1.611(2)$ \cite{Tilley1993,Ajzenberg1991}, $1.585(3)$ \cite{Tilley1995,Ueno2013}, and $1.84(9)$ \cite{Stanoiu2004} MeV, respectively.
It is worthwhile noting that the lowering of the neutron $s_{1/2}$ state relative to $d_{5/2}$ state as the
dripline is approached is expected for a simple potential \cite{Millener2001},
and is further enhanced by the effects of weak binding \cite{Tanihata1996} as argued for by Ref.~\cite{Hoffman2014}.

Finally it is interesting to observe that 
the $5/2_1^+$ states in \nuc{19}{C} and \nuc{23}{O} \cite{Schiller2007,Tshoo2014} (both $T_z = 7/2$) are each narrow resonances lying only around 50~keV above the neutron decay threshold.
This is somewhat surprising as \nuc{23}{O} 
has a deeper neutron binding potential well -- $E_x(5/2_1^+) - E_x(1/2_1^+)$ $\approx$ 2.8~MeV.
Whether such behavior is a coincidence or has an underlying explanation would be interesting to investigate further. 


In conclusion, single-neutron  
knockout from $^{20}\textrm{C}$ has been measured at 
280~MeV/nucleon and three unbound levels observed in $^{19}\textrm{C}$. 
Hole states -- $J^{\pi}=5/2^+$ and $1/2^-$ -- created by  
removing neutrons from the $0d_{5/2}$ and $0p_{1/2}$ orbits were populated and 
identified by the associated longitudinal momentum distributions.
Comparison with eikonal-model reaction calculations permitted spectroscopic factors to
be deduced which were found to be in reasonable accord with shell-model calculations.
The large specroscopic strength observed for the population of the $5/2^+_1$ state indicates that the $^{20}\textrm{C}$ ground state
valence neutron configuration includes, in addition to the known $1s_{1/2}^2$ 
component, a significant $0d_{5/2}^2$ contribution.
In terms of the level scheme of $^{19}\textrm{C}$, the YSOX interaction, developed from the monopole-based
universal interaction, provided the best description, 
including, most notably, the energy of the newly observed $1/2^-$ cross-shell state.  In this context, determining the location of the corresponding  $3/2^-$ level, which would appear to lie higher in excitation energy, would be of considerable interest.

\section*{Acknowledgements}

We are grateful to Dr.~C.~Yuan
for providing us with their shell-model interaction,
and to Prof.~J.~A.~Tostevin for fruitful discussions.
The work presented here was in part supported by 
the WCU (R32-2008-000-10155-0) and
the GPF (NRF-2011-0006492) programs of NRF Korea,
JSPS KAKENHI Grant Number 16H02179
and MEXT KAKENHI Grant Number 24105005 in Japan, and
the French ANR grant EXPAND (ANR-14-CE33-0022-01).
NLA, FD, JG, FMM, and NAO acknowledge
partial support from the French-Japanese LIA-International Associated Laboratory
for Nuclear Structure Problems.
AN and JG would like to acknowledge the JSPS Invitation fellowship program
for long term research in Japan at the Tokyo Institute of Technology and RIKEN, respectively.
SL gratefully acknowledges the support provided by the RIKEN International Associate Program
and the hospitality of the Nishina Center Staff during his sejour.


\bibliography{hwang_19c}

\end{document}